\documentclass[aps,showpacs,preprintnumbers,amsmath,amssymb,twocolumn,superscriptaddress,prl]{revtex4-1}
\usepackage{float}
\usepackage{graphicx}% Include figure files
\usepackage{dcolumn}% Align table columns on decimal point
\usepackage{bm}% bold math
\usepackage{color}
\usepackage{subfigure}  % use for side-by-side figures
\def\comment#1{}

\def\slashchar#1{\setbox0=\hbox{$#1$}           % set a box for #1
	\dimen0=\wd0                                 % and get its size
	\setbox1=\hbox{/} \dimen1=\wd1               % get size of /
	\ifdim\dimen0>\dimen1                        % #1 is bigger
	\rlap{\hbox to \dimen0{\hfil/\hfil}}      % so center / in box
	#1                                        % and print #1
	\else                                        % / is bigger
	\rlap{\hbox to \dimen1{\hfil$#1$\hfil}}   % so center #1
	/                                         % and print /
	\fi}                                         %

\begin{document}

\title{First-principles analysis of nanoelectromechanical systems using Loewner equation}

\author{Edgar Marcelino}
%\email{edgarufba@gmail.com}
\address{Centro Brasileiro de Pesquisas F\'{i}sicas, Rua Dr. Xavier Sigaud 150, 22290-180, Rio de Janeiro, RJ, Brazil}

\author{Thiago A. de Assis}
%\email{thiagoaa@ufba.br}
\address{Instituto de F\'{\i}sica, Universidade Federal da Bahia,
   Campus Universit\'{a}rio da Federa\c c\~ao,
   Rua Bar\~{a}o de Jeremoabo s/n,
40170-115, Salvador, BA, Brazil}

\author{Caio M. C. de Castilho}
%\email{caio@ufba.br}
\address{Instituto de F\'{\i}sica, Universidade Federal da Bahia,
   Campus Universit\'{a}rio da Federa\c c\~ao,
   Rua Bar\~{a}o de Jeremoabo s/n,
40170-115, Salvador, BA, Brazil}
\address{Instituto Nacional de Ci\^{e}ncia e Tecnologia em Energia e Ambiente - INCTE\&A, Universidade Federal da Bahia,
   Campus Universit\'{a}rio da Federa\c c\~ao,
   Rua Bar\~{a}o de Jeremoabo s/n,
40170-280, Salvador, BA, Brazil}
\address{Centro Interdisciplinar em Energia e Ambiente, Universidade Federal da Bahia,
   Campus Universit\'{a}rio da Federa\c c\~ao, 40170-115, Salvador, BA, Brazil}

\author{Roberto F. S. Andrade}
%\email{randrade@ufba.br}
\address{Instituto de F\'{\i}sica, Universidade Federal da Bahia,
   Campus Universit\'{a}rio da Federa\c c\~ao,
   Rua Bar\~{a}o de Jeremoabo s/n,
40170-115, Salvador, BA, Brazil}

\begin{abstract}
The Loewner equation (LE) is used to obtain conformal mappings that lead to exact and analytical expressions for several electrostatic properties of realistic quasi-unidimensional nano-electromechanical systems (NEMS). The LE approach also embraces curved geometries, impossible to be addressed by traditional methods such as the Schwarz-Christoffel transformation, often used in this scenario. Among the possible applications of the formalism, we show that it allows for an exact evaluation of the field enhancement factor (FEF) close to the apex of different emitters. Despite its key role in the demodulation process for radio-receiver nano-devices, actual FEF values have been mostly obtained via numerical and/or phenomenological approaches. This work extends the already huge universe of applications of the LE and provides an analytical method to evaluate the FEF, even for curved emitters. Furthermore, our results provide a signature of the varying emitted current's response due to the nanostructure oscillation, justifying its role in the demodulation process of radio-frequency.
\end{abstract}

\maketitle

The study of nanoelectromechanical systems (NEMS) \cite{NEMS,Nature2004} currently attracts great attention of the scientific community, not only due to the interesting theoretical aspects involved \cite{Gusso,Schmidt,Metelmann,Gorelik}, but also as a result of the enormous number of potential applications that can be derived among the many issues related to the field. Some examples include quantum nanomechanical resonators \cite{LaHaye}, single-molecule detection \cite{YangNanoLett}, chemical, mass and thermal sensing \cite{Wan,Waggoner,Roukes,Zettl,PRAP2018}, integrated circuits \cite{Feng}, high-frequency signal sources/generation \cite{Feng,Unterreithmeier,Mahboob}  and field emitting nanotubes operating similarly to diode detectors \cite{Zettl2}. Most of these applications involve the oscillation of nanotubes \cite{Cole2015chapter}, or other similarly shaped structures, with lateral dimensions around a few nanometers \cite{NEMS}. Besides that, NEMSs formed by field emission (FE) diode-like nano-detectors \cite{PRB2011,Purcell} must present a large aspect ratio.

 Under the action of an external macroscopic electrostatic field, $\mathbf{E_0}$, the local field close to the apex of a NEMS is largely enhanced, which is measured by the field enhancement factor (FEF), reaching typical values  $\sim \,10^{2}- 10^{3}$. This makes carbon nanotubes (CNTs) \cite{FENanotube,Cole2015chapter} suitable for producing related technological applications \cite{Nature2004,CondMatter, Zettl2, Ayary,Zettl3}. Indeed, by field-induced emission of electrons, it was possible to control the resonance vibrations of CNTs with 40 $\mu$m height and radii in the range between 10 and 20 nm (aspect ratio $\sim 10^{3}$), when the tip anode is a few millimeters far away from the nanotube apex, under ultra high vacuum conditions \cite{Purcell}. For such conditions, the nanotube apex-FEF, which is evaluated at a well characterized distance from the uppermost atom as we will define latter, is expected to depend only on the geometry \cite{Forbes2012b,deAssisAPL2018}. These limits, which are valid for technological purposes, will be taken into account here.

 When a CNT is excited by a Lorentz force, the spatial and temporal variations of its apex-FEF during oscillation becomes a key parameter for the demodulation process in radio-frequency NEMS \cite{Zettl2,PRB2011,Nature2009}. This occurs because the response of the emitted current to the nanotube oscillation is proportional to the time variation of the apex-FEF \cite{Zettl2,PRB2011}. On the other hand, knowing the apex-FEF of an oscillating NEMS is also essential to produce sensitive nano-detectors of adsorbed atoms. In this case, the resonance frequency of a NEMS, formed by a CNT, is shifted even when a single atom is adsorbed on its cap, since it is sensitive to the apex local field \cite{Purcell}. Thus, resonators with a high quality factor can be used as potential detectors of single adatoms. CNTs have already proved to exhibit high quality mechanical factors at low temperatures \cite{NNCNTB,PRL2018}, while similar properties of metallic nanowires have been recently uncovered \cite{PRL20182}.

 In opposition to the experimental advances, precise analytical results for the apex-FEF of quasi-unidimensional oscillating emitters are still lacking, constituting a highly non-trivial challenge to support NEMS technologies. Despite some limited numerical results \cite{PRB2011}, its evaluation in a large number of works is often neglected or subject of rough estimates \cite{CondMatter,Zettl2}. In principle, it requires to consider the classical representation of a CNT/nanowire emitter with the shape of a deformed hemisphere on a cylindrical post (HCP) model \cite{Ultramicroscopy,Xanthakis,RBowring,ZENG2009,deAssis1}, or any equivalent counterpart in smaller dimensions. In this letter, we address this problem by taking into account the large aspect-ratio of the nanotube (or other nanostructure). This justifies modeling it by a line,  and solving analytically a two-dimensional problem of a deformed nanostructure starting on an infinite conducting line. We obtain a suitable conformal mapping to perform a first-principles evaluation of the FEF in the vicinity of the tip of the nanostructure,  where the charge density is maximal. Since actual nanostructures are modeled by possibly curved unidimensional lines, standard conformal mapping techniques \cite{Miller1,Miller2,Qin,Qin2,Tang,Venkattraman,MarcelinoJVSTB,MarcelinoJAP}, as the Schwarz-Christoffel transformation (SCT) \cite{SC1,SC2}, are not suitable to solve the problem.  Therefore, we consider the Loewner's equation (LE) approach \cite{Loewner} to obtain the desired conformal mappings. Our results show the clear advantages of the method in obtaining exact results for realistic systems.

\begin{figure}[H]
\centering    % figura centralizada
\includegraphics[width=0.27\textwidth]{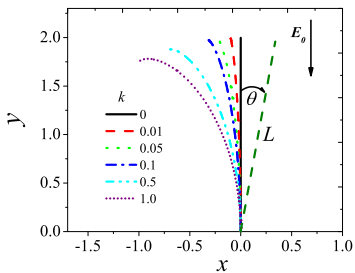}
\vspace{-0.4cm}
\caption{(Color Online) Lines of singularities originated by driving functions $\xi_i(t)$, $i=1-3$, used to model nanowires: i) $\xi_1$, vertical slit (solid black line); ii) $\xi_2$, tilted straight line making an angle $\theta$ with the vertical direction (dark green dashed line); iii) $\xi_3$, curved lines with the shape of logarithmic spirals, corresponding to different values of $k$ in Eq.(\ref{spiral3}): $0.01$ (red dashes), $0.05$ (green short dashes), $0.1$ (blue dash-dot), $0.5$ (cyan  dash-dot-dot) and $1.0$ (wine dots).}	 \label{NeedleEmitter}
\end{figure}

Being originally proposed \cite{Loewner} to study the Bieberbach's Conjecture \cite{Bieberbach1,Bieberbach2,Branges}, the LE approach plays now an important role in many different areas such as Laplacian growth, fractals, conformal field theories (CFT), random walks, and percolation, among many others. The LE approach led to a reformulation of the diffusion limited aggregation (DLA) model \cite{Witten,Hastings} through a deterministic version \cite{Carleson}, what has motivated other works using LE in surface growth processes \cite{Gubiec}. Despite the impact of  early works, a revolution in this area began by noticing the connection between a one-dimensional Brownian motion and a  fractal growth in a plane, which inspired the development of a new framework: the Schramm-Loewner evolution (SLE) \cite{Schramm1,Schramm2,Schramm3}.

The (chordal) LE allows to obtain a conformal transformation $g: \cal{H} \rightarrow \mathcal{H}$$/ \Gamma_{t}$ that maps the upper complex half-plane ($\cal{H}$) into itself minus a set of lines ($\Gamma_{t}$), which are Jordan arcs. For a single arc, the LE reduces to
\begin{equation}   \label{Loewner}
\frac{\partial g(z,t)}{\partial t}=\frac{2}{g(z,t)-\xi(t)}.
\end{equation}
For each value of $t$, the function $w=g(z,t)$ maps the upper complex half-plane minus a line into the whole upper complex half-plane. As $t$ increases, the line removed from the original plane evolves, making a slit in the complex domain. The real valued ``driving function", $\xi(t)$, determines the shape of the line of singularities $z(t)=z_{c}(t)$, corresponding to the aforementioned slit, and is obtained from the condition $g(z_{c}(t),t)=\xi(t)$. A solution of Eq. (\ref{Loewner}) determines the conformal mapping with the desired properties, provided it obeys the initial (\ref{Initial}) and the hydrodynamic (\ref{Hydrodynamic}) conditions:
\begin{eqnarray}
g(z,t=0)=z,     \label{Initial}  \\
|z| \rightarrow \infty \Rightarrow g(z,t)=z+O \left( \frac{1}{|z|} \right). \label{Hydrodynamic}
\end{eqnarray}
\noindent  Although some families of solutions of the LE are well known \cite{Kadanoff2},  there is no general method either to exactly solve it for an arbitrary $\xi(t)$, or to determine the function $\xi(t)$ that generates a given line of singularities \cite{Kadanoff1}.

Here we model a conducting nanowire on an infinity conducting line in a plane, under an intense electrostatic field. We consider singularity lines
such that: i) the solutions $z_{c}(t)$ can be analytically evaluated; ii) the shape of $z_{c}(t)$ reproduces a  nanostructure bending sidewards during NEMS oscillation. With these assumptions, we evaluate realistic FEF values ($\gamma$) close to NEMS tips, based on the following considerations: i) At any given instant $t$, $g(z,t)$ maps the upper complex half-plane $z=(x,y) \rightarrow x+iy$ minus the line of singularities into the upper complex half-plane $w=(u,v) \rightarrow u+iv$; ii) The electrostatic fields in the $z$ and $w$-planes are respectively given by: $E_{x}-iE_{y}=\partial \phi / \partial z=(\partial \phi /\partial w)(\partial w/ \partial z)$ and $E_{u}-iE_{v}=\partial \phi / \partial w$.
Therefore, a general expression can be derived for the FEF as
\begin{equation}\label{eqfef}
\gamma(x,y) \equiv \frac{\left|\mathbf{E(x,y)}\right|}{\left|\mathbf{E_0}\right|}=\left|\frac{dz}{dw} \right|^{-1}.
\end{equation}

We model our system using three functions $\xi_i(t)$,
\begin{equation}\label{eqxi}
\xi_1(t)=A, \, \xi_2(t)=2\sqrt{kt}, \,\xi_3(t)=2\sqrt{k(1-t)},
\end{equation}
where $A$ and $k$ are constants. All of them lead to exact solutions of Eq. (\ref{Loewner}), with lines of singularities corresponding respectively to a vertical straight slit, an oblique straight slit or a logarithmic spiral \cite{Kadanoff2} (see Fig. \ref{NeedleEmitter}). Therefore, starting from the symmetric case ($\xi_1(t)$), the nanostructure may remain a straight line changing the angle $\theta$ with the vertical line ($\xi_2(t)$), or be subject to deformation, changing then its curvature ($\xi_3(t)$).

\begin{figure}[H]
\centering    % figura centralizada
\includegraphics[width=0.26\textwidth]{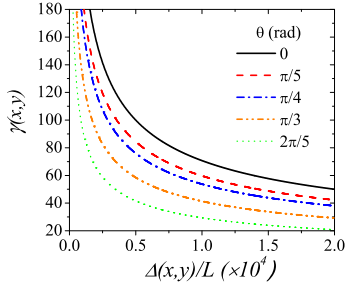}
\vspace{-0.4cm}
\caption{(Color Online) Dependence of $\gamma$, for straight tilted nanowires originated by $\xi_2$ (see Fig. \ref{NeedleEmitter}), on $\Delta(x,y)/L$ for different values of $\theta$.}	\label{NeedleFEFd}
\end{figure}

For $\xi_1(t)$, we can easily perform a direct integration of Eq. (\ref{Loewner}), satisfying Eqs. (\ref{Initial}-\ref{Hydrodynamic}), obtaining then $g(z,t)=A + \sqrt{(z-A)^{2}+4t}$.  The line of singularities in the $z$-plane is given by $z_{c}(t)=A+2i\sqrt{t}$. Introducing a parametrization $t=t^*_1=L^{2}/4$, where $L$ corresponds to the size of the nanostructure, we see that $g(z,t)$ maps the upper complex half-plane $z=(x,y)$ minus the vertical slit starting at $z=A$ and ending at $z=A+Li$, into the upper complex half-plane $w=(u,v)$. The desired conformal mapping is obtained by inverting $w=g(z(w),t^*_1)$, which leads to $z=A+\sqrt{(w-A)^{2}-L^{2}}$. When $w\approx A$, we can expand $\sqrt{(w-A)^{2}-L^{2}} \approx iL$. Without loss of generality, for an infinity conducting line we can assume that the slit starts at $z=0 \Leftrightarrow A=0$, so that Eq. (\ref{eqfef}) leads to
\begin{equation} \label{FEFI}
\gamma(x,y) \approx \sqrt{\frac{L}{2|z-iL|}},
\end{equation}
where $|z-iL|=\sqrt{x^{2}+(y-L)^{2}}$ is the distance between the point $(x,y)$, where the FEF is evaluated, and the tip. As the applied field $\mathbf{E_0}$ is known and the direction of the field is perpendicular to the conducting emitter, the local electric field results being completely determined by $\gamma$. The charge density may also be immediately derived, since it can be obtained from the electric field by Gauss Law. Fig. \ref{NeedleFEFd} illustrates the behavior of $\gamma$ close to the tip of a vertical conducting slit of size $L$, perpendicular to an infinity conducting line ($\theta=0$), according to Eq.(\ref{FEFI}).

The vertical slit can be viewed as the limit of an infinity line with an isosceles triangular protrusion of height $L$ and half-width $a$, when $a$ tends to zero. The evaluation of $\gamma$ near the apex ($z=iL$) of this system \cite{MarcelinoJVSTB} results in
\begin{equation} \label{FEFISO}
\gamma \approx \left[\frac{\sqrt{\pi(a^{2}+L^{2})}}{(2-\alpha) \Gamma \left(1-\frac{\alpha}{2} \right) \Gamma \left(\frac{1+\alpha}{2} \right) |z-iL|} \right]^{(1-\alpha)/(2-\alpha)},
\end{equation}
where $\alpha \equiv \frac{2}{\pi} \arctan \left( \frac{a}{L} \right)$. As can be seen, Eq. (\ref{FEFI}) is recovered from Eq.(\ref{FEFISO}) in the limit $a \rightarrow 0 \Rightarrow \alpha \rightarrow 0$.

The solution of Eq. (\ref{Loewner}), satisfying Eqs. (\ref{Initial}-\ref{Hydrodynamic}) for the driving function $\xi_2(t)$, can be found in \cite{Kadanoff2}. After some manipulation on the reported results, we arrive at
\begin{equation} \label{Conformal2}
[z(w)]^{y_{+}-y_{-}}=\frac{(g-y_{-} \sqrt{t})^{y_{+}}}{(g-y_{+} \sqrt{t})^{y_{-}}},
\end{equation}
where $y_{\pm} \equiv \sqrt{k} \pm \sqrt{k+4}$. The corresponding line of singularities \cite{Kadanoff2} reads
\begin{equation} \label{SingLine2}
z_{c}(t)=2\sqrt{t} \left[\frac{\sqrt{k+4}+\sqrt{k}}{\sqrt{k+4}-\sqrt{k}} \right]^{\frac{\sqrt{k}}{2 \sqrt{k+4}}} e^{\frac{i \pi}{2} \left(1-\frac{\sqrt{k}}{\sqrt{k+4}}\right)}.
\end{equation}
Introducing $t=t^*_2=L^{2}/4 \left[\frac{\sqrt{k+4}+\sqrt{k}}{\sqrt{k+4}-\sqrt{k}} \right]^{\frac{\sqrt{k}}{\sqrt{k+4}}}$ and inverting $w=g(z(w),t^*_2)$, we get a conformal transformation mapping the upper complex half-plane $w$ into the upper complex half-plane $z$, minus a straight slit of size $L$ forming an angle $\theta(k)=\frac{\pi}{2} \frac{\sqrt{k}}{\sqrt{k+4}}$ with the vertical axis (see Fig. \ref{NeedleEmitter}). When $w=\xi_2(t^*_2)$, we obtain the tip coordinate $z=z_{c}(t^*_2)$. Thus, Eq. (\ref{Conformal2}) can be rewritten as
 \begin{equation}  \label{Needle1}
 z=z_{c} \left[1+ \frac{w-2\sqrt{kt}}{y_{+} \sqrt{t}} \right]^{\frac{1}{2}+\frac{\sqrt{k}}{2\sqrt{k+4}}}  \left[1+ \frac{w-2\sqrt{kt}}{y_{-} \sqrt{t}} \right]^{\frac{1}{2}-\frac{\sqrt{k}}{2 \sqrt{k+4}}}.
\end{equation}
Near the apex ($w \approx 2\sqrt{kt}$), Eq. (\ref{Needle1}) and its derivative with respect to $w$ lead to the approximations:
 \begin{eqnarray}
 |z-z_c| \approx \frac{|w-2\sqrt{kt}|^{2}}{8t}  |z_c|,   \label{Needle3}   \\
  \left|\frac{dz}{dw} \right|=\frac{|z_c||w-2\sqrt{kt}|}{4t}.   \label{Needle4}
 \end{eqnarray}
Using Eq. (\ref{eqfef}), $t^*_2$, and $\theta(k)$, Eqs. (\ref{Needle3}) and (\ref{Needle4}) lead to
 \begin{equation}  \label{FEFII}
\gamma(x,y)=\left( \frac{\pi-2 \theta}{\pi+2 \theta}\right)^{\theta/\pi} \sqrt{\frac{L}{2 \Delta }}.
\end{equation}

Here, the distance from $(x,y)$ to the tip is given by $\Delta=| z-Le^{i\left(\frac{\pi}{2}-\theta\right)} | =\sqrt{(x-L \sin \theta)^{2}+(y-L \cos \theta)^{2}}$. As expected, for $\theta=0$ the slit is vertical and Eq. (\ref{FEFI}) is recovered. In Fig. \ref{NeedleFEFd}, $\gamma(x,y)$ is plotted as a function of the distance to the apex for different angles. One can see that $\gamma$ monotonically decreases with the distance to the tip and also with $\theta$.
It is worthy noting that the FEF close to the apex becomes as large as $10^{2}$ for $\Delta/L \approx 10^{-5}$ when $\theta=0$, which lies well in the range of experimental results. In fact, the apex-FEF is rigorously defined at the position, on the vacuum side, of the geometrical plane of the outermost surface atom nuclei, defining then the repulsion distance \cite{ForbesES}. This distance is comparable with the atomic radius.  Hereafter we define $\gamma_{ap}$ as the apex-FEF, i.e., the value of $\gamma$ at the immediate apex vicinity of the structure, for which we consider $\Delta/L = 10^{-5}$. Thus, our results provide a reasonably faithful description for a NEMS consisting of a nanotube with $L=10 \mu$m, where $\Delta \approx 1 \AA$ is of the order of the carbon atomic radius. Moreover, recent results have shown that, for distances of this same order, Density Functional Theory (DFT) results for the apex-FEF agree well with the classical ones \cite{Lepetit2016}. In the inset of Fig. \ref{NeedleFEFthetaSpiral}, $\gamma_{ap}$ is plotted as a function of the angle $\theta$.

Finally, we consider the case of a curved nanowire, which can be obtained after inserting $\xi_3(t)$ into Eq. ({\ref{Loewner}). In this case, it is possible to show that the conformal mapping and the line of singularities obey the following expressions:
\begin{eqnarray}\label{spiral}
\frac{(z-y_{-})^{y_+}}{(z-y_{+})^{y_-}}=\frac{(w-y_{-}\sqrt{1-t})^{y_+}}{(w-y_{+}\sqrt{1-t})^{y_-}},  \\
\frac{(z_{c}-y_{-})^{y_+}}{(z_{c}-y_{+})^{y_-}}=\frac{y_{+}^{y_+}}{y_{-}^{y_-}}(1-t)^{(y_{+}-y_-)/2},
\end{eqnarray}
where now $y_{\pm}=\sqrt{k}\pm i\sqrt{4-k}$ \cite{Kadanoff2}. The line of singularities corresponds to a logarithmic spiral, as illustrated in Fig. \ref{NeedleEmitter}. We have considered only cases where $k<1$, in which the tip of the line of singularities never points downwards. Proceeding similarly to the previous case, Eq. ({\ref{spiral}}) may be rewritten as:
\begin{equation}\label{spiral2}
\frac{\left[1+\frac{w-2\sqrt{k(1-t)}}{y_{+} \sqrt{1-t}} \right]^{y_+} }{\left[1+\frac{w-2\sqrt{k(1-t)}}{y_{-} \sqrt{1-t}} \right]^{y_-}}=\frac{\left[1+\frac{z-z_c}{z_{c}-y_{-}} \right]^{y_+} }{\left[1+\frac{z-z_c}{z_{c}-y_{+}} \right]^{y_-} }.
\end{equation}
Following the general result in Eq. (\ref{eqfef}) and the steps used in the case of the tilted slit, Eq. (\ref{spiral2}) is first derived with respect to $z$. After somewhat lengthy calculations, the resulting expressions in the limit $w\rightarrow 2\sqrt{k(1-t)} \Rightarrow z \rightarrow z_c$ lead to the desired expression for $\gamma$:
\begin{figure}[H]
\centering    % figura centralizada
\includegraphics[width=0.27\textwidth]{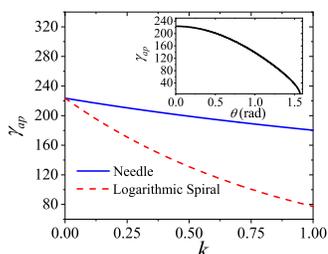}
\vspace{-0.6cm}
\caption{(Color Online) Dependence of $\gamma_{ap}$ on the parameter $k$ for lines of singularities obtained with $\xi_2$ (solid blue line) and $\xi_3$ (red dashed  line). Results correspond to $\Delta(x,y)/L=10^{-5}$ and $t=0.9999$ for the logaritmic spiral. $\Delta(x,y)$ is the distance to the tip. The inset highlights the dependence of $\gamma_{ap}$ on $0\le\theta\lesssim\pi/2$, since the upper limit corresponds to $k\rightarrow \infty$.} \label{NeedleFEFthetaSpiral}
\vspace{-1cm}
\end{figure}
%\begin{widetext}
\begin{equation}
\gamma(x,y)=\frac{\sqrt{(1-t)|z_{c}(t)(y_{+}-y_{-})+y_{-}^{2}-y_{+}^{2}|}}{\sqrt[4]{4-k}\sqrt{|z_{c}(t)-y_{-}||z_{c}(t)-y_{+}||z-z_c(t)|}}.
\end{equation}
It is possible to prove that $t \rightarrow 1^{-} \Rightarrow z_{c}(t) \rightarrow y_+$. In this limit, the asymptotic expression for $|z_{c}(t)-y_{+}|$ reads
\begin{equation}\label{spiral3}
|z_{c}(t)-y_{+}| \approx \left [\frac{16(4-k)^{k-2} (1-t)^{4-k}}{e^{-2\zeta \sqrt{k(4-k)}}} \right ]^{1/4} ,
\end{equation}
%\end{widetext}
where $\zeta=\arctan \left[ \sqrt{4-k}/\sqrt{k} \right] -\pi /2$. Finally, the derivation of the proper expression for $\gamma$ still requires a parametrization relating $t$ with the size of the emitter. The situation is different from the one of an oblique slit, since the curve is spiralling towards $y_+$ for $t \rightarrow 1^{-}$. The parameter $0<k<1$ controls the curvature and, for $k=0$, the spiral becomes a vertical slit of size 2. In order to solve this limitation, we introduce another conformal transformation that provides a simple isotropic dilation. Therefore, we are led to the final expression for $\gamma$ close to the tip $z_{c}=y_{+}$ ($t \rightarrow 1^{-}$):
\begin{equation}  \label{FEFIII}
\gamma(x,y)=\frac{(1-t)^{k/8}}{(4-k)^{k/8}} e^{-\zeta \sqrt{k(4-k)}/4} \sqrt{\frac{L}{2|z-\frac{Ly_{+}}{2}|}}.
\end{equation}
The parameter $L$ controls the length of the spiral for different $k$ and can be determined by performing an integration over arc-length parametrization. It corresponds to the size of the slit for $k=0$. Figure \ref{NeedleFEFthetaSpiral} shows the behavior of $\gamma_{ap}$ for the emitters in Fig. \ref{NeedleEmitter} as a function of $k$. As expected, for $k \rightarrow 0$ the results collapse to those in Eq. (\ref{FEFI}). The effect of the curvature, which provides a more realistic morphology of a nanowire during oscillations, is to strongly reduce the value of $\gamma_{ap}$ as compared to the needle shaped case. This analytical result provides a strong evidence supporting the amplified nature of the emitted current's response to the nanotube oscillation, justifying its role in the demodulation process for radio-frequency NEMS. This opens precedent to further explorations of LE as a powerful technique to understand NEMS properties.

Summarizing, in this work we have demonstrated how the LE can be used to evaluate the FEF of quasi-unidimensional geometries. This method improves the study of emission properties for devices with very large aspect-ratio (possibly nanotubes/nanowires) used in NEMS. We have shown that, with a proper choice of the driving function $\xi(t)$, it is possible to model different NEMS morphologies, like a tilted straight rod or a curved device, which are actually revealed in transmission electron microscopy studies. Oscillatory behavior of nanostructures can also be modeled by letting $k$ be time-dependent. %The shape of the nanowire during an oscillation is approximated by a logarithmic spiral and, in the case of a straight line, we show that the results obtained via SCT are recovered.
The present work expands the already huge universe of applications of the LE, by applying it to NEMS formed by single tip field emitters. We hope our analytical results help to elucidate the connection between the apex-FEF and the demodulation processes in radio-receiver NEMS, for quasi-unidimensional field emitter devices. Other relevant problems that may be treated with this approach (from analytical and/or numerical perspective) include the electric field depolarization due to proximity between emitters and the evolution of the emitted current, obtained via Fowler-Nordheim-type equation, during stochastic surface growth in multiple tip NEMS.

\vspace{-0.1cm}

This study was financed in part by the Coordena\c{c}\~{a}o de Aperfei\c{c}oamento de Pessoal de N\'{i}vel Superior - Brasil (CAPES) - Finance Code 001. The authors also thank CNPq (Brazilian Agency). 
%Edgar Marcelino acknowledges financial support from CAPES.

%\bibliography{SCbib}


\end{document}